
\baselineskip=15pt
\vsize=19.3cm
\hsize=13.4cm
\def\part{\partial}

\def\pmb#1{\setbox0=\hbox{#1}\kern-.025em
    \copy0\kern-\wd0\kern.05em\kern-.025em\raise.029em\box0}

\def\a{\alpha}      \def\l{\lambda}   \def\L{\Lambda}
\def\b{\beta}       \def\m{\mu}
\def\g{\gamma}      \def\G{\Gamma}    \def\n{\nu}
\def\d{\delta}      \def\D{\Delta}         
\def\e{\epsilon}                      
\def\ve{\varepsilon}                  \def\s{\sigma}  
                         
        \def\vphi{\varphi}
                   \def\o{\omega}

\def\cF{{\cal F}}

\def\cL{{\cal L}}

\def\fr#1 #2{\hbox{${#1\over #2}$}}        
\def\leaderfill{\leaders\hbox to 1em{\hss.\hss}\hfill}

\def\section#1{                            
\vskip.6cm\goodbreak                       
\noindent{\bf \uppercase{#1}}
\nobreak\vskip.4cm\nobreak  }

\def\subsection#1{
  \vskip.4cm\goodbreak
  \noindent{\bf #1}
  \vskip.3cm\nobreak}

\def\subsub#1{\par\vskip4pt {\bf #1} }

\def\ver#1{\left\vert\vbox to #1mm{}\right.}

\font\eightrm=cmr8                                
\font\tfont=cmbx12                                

\def\preprint#1{\hskip10cm #1 \par}
\def\date#1{\hskip10cm #1 \vskip2cm}
\def\title#1{ \centerline{\tfont{#1}} }
\def\titlef#1{ \vskip.2cm                          

      \centerline{ \tfont{#1}
                   \hskip-5pt ${\phantom{\ver{3}}}^\star$  }
      \vfootnote{$^\star$}{\eightrm Work supported in part
      by the Serbian Research Foundation, Yugoslavia.} \vskip.5cm }
\def\author#1{ \vskip.5cm \centerline{#1} }
\def\institution#1{ \centerline{\it #1} }

\def\abstract#1{ \vskip3cm \centerline{\bf Abstract}
                 \vskip.3cm {#1} \vfill\eject }


\magnification=1200
\vskip1cm

\preprint{IF--S5/95}
\date{March, 1995}
\title{\bf BRST ANALYSIS OF GAUGE THEORIES BASED ON}
\titlef{\bf NONLINEAR ALGEBRAS IN $2d$ }
\author{M. Blagojevi\'c}
\institution{Institute of Physics,  P.O.Box 57, 11001 Beograd, Yugoslavia}
\author{T. Vuka\v sinac}
\institution{Department of Theoretical Physics, Institute of Nuclear
Sciences, Vin\v ca,}
\institution{P.O.Box 522, 11000 Beograd, Yugoslavia}

\abstract{Covariant quantization of theories based on nonlinear
extensions of Lie algebras in $2d$ is studied by using a generalized
Lagrangian BRST formalism. The quantum action is constructed to be
invariant under the off--shell nilpotent BRST transformations by
using a set of independent antifields as auxiliary, nonpropagating
variables in the quantum theory. The general results are applied to the
quantization of nonlinear gauge theory based on quadratic Poincar\'e
algebra, which is closely related to $2d$ gravity with dynamical torsion. }

\subsection{1. Introduction} 

General relativity is a succsessful  theory of macroscopic gravitational
phenomena, but all attempts to quantize the theory encounter serious
difficulties. It seems natural to try to understand the structure of
gravity on the basis of the concept of gauge symmetry, which has been
very successful in describing other fundamental interactions in nature.
The usual gauge theory is based on a symmetry group with local
properties determined by a Lie algebra. Ikeda and Izawa [1,2]
considered an interesting approach to the construction of gauge
theories in two dimensions, based on a {\it nonlinear extension} of the
Lie algebra structure. It turns out that the resulting nonlinear gauge
theory is closely related to two\---dimensional gravity. In case when
the nonlinear algebra is a quadratic extension of the Poincar\'e algebra,
they obtained a gauge theory which is, in a certain region, equivalent
to two\---dimensional gravity with dynamical torsion [1,3], or to
dilaton gravity [2,3]. In a similar way, the quadratic $W_3$ algebra is
shown to lead to the $W_3$ theory of gravity [4,5].

One expects that investigations of two\---dimensional theories of
gravity may provide a better understanding of quantum properties of
higher dimensional gravity, as well as a deeper understanding of string
theory. For this reason nonlinear gauge theories may be of importance in
considerations related to quantum gravity.

In this paper we shall study the BRST quantization of theories based on
nonlinear extensions of Lie algebras in $2d$. Ikeda and Izawa described
the construction of the action for nonlinear gauge theory, whose
classical gauge symmetries are characterized by a gauge algebra which
is closed only {\it on shell} [1-3]. The gauge theories of this type
can be covariantly quantized by using the general Lagrangian method
developed by Batalin and Vilkovisky (BV) [6]. The method is based on a
generalization of the BRST approach, but the effective BRST
transformations obtained in the process of gauge fixing are nilpotent
only {\it on shell}.  The BRST analysis of gauge theories based on the
nonlinear Poincar\'e algebra and the $W_3$ algebra have been done in
references [1] and [5], respectively.  In both cases the gauge--fixed
theory is characterized by an on--shell nilpotent BRST symmetry.  On
the other hand, the generalization of the BV method, which yields an
{\it off shell} nilpotent BRST symmetry of the gauge--fixed action, has
been proposed in Ref. [7]. It has been successfully used to quantize
Witten's interacting bosonic string field theory, the superparticle in
$d=10$ and $d=9$, the heterotic superstring and the simple supergravity
[8]. In this approach, the gauge--fixed theory is realized on a set of
fields containing a convenient set of antifields. The elimination of
the antifields leads to the BV form of the theory, showing clearly that
the essential role of the antifields is to ensure the off--shell
nilpotency of the BRST transformations in the gauge--fixed action.

We shall begin our exposition in sect. 2 by considering the form of the
classical gauge structure of the {\it general} nonlinear gauge theory
[1-3]. The basic features of the Lagrangian gauge algebra are
characterized by a set of structure functions, determined by the
Poisson bracket algebra and the Jacobi identities of the gauge
generators. In sect. 3 we find the form of the BRST generator as the
solution of the master equation, by making use of the classical
structure functions. Then, we find a BRST invariant extension of the
classical action and define the quantum theory by an appropriate
gauge--fixing procedure.  The quantum action is constructed to be
invariant under the {\it off--shell} nilpotent BRST transformations,
which is achieved by introducing a set of antifields as auxiliary,
nonpropagating variables in the quantum theory. The relation to the BV
approach is clarified by observing that the elimination of the
antifields yields an effective theory which coincides with the BV form.
In sect. 4 we apply the general results of the previous section to
quantize the gauge theory based on the {\it quadratic} extension of the
Poincar\'e algebra.  Section 5 is devoted to conclusions.

\subsection{2. Classical theory and gauge symmetry} 

Let us consider a nonlinear extension of a Lie algebra with a basis
$T_a$, the bracket product of which is given by the relation
$$
[T_a,T_b]=W_{ab}(T)  \, ,                                     \eqno(1a)
$$
where $W_{ab}(T)=-W_{ba}(T)$ is an antisymmetric polinomial in $T_c$,
$$
W_{ab}(T)=k_{ab}+f_{ab}{^c}T_c +V_{ab}^{cd}T_cT_d +\cdots\, ,
$$
and the coefficients of $k$, $f$, $V$, and so on, are the structure
{\it constants}  of the algebra. One also assumes that the bracket
product is a derivation, eg. $[A,BC]=[A,B]C+B[A,C]$.
The zeroth order term $k_{ab}$ is the central element of the algebra,
the first order term $f_{ab}{^c}$ is the usual structure constant of
the corresponding Lie algebra, and all other terms characterize
nonlinear structure of the algebra. The antisymmetry of $W_{ab}(T)$
implies the following symmetry properties of the structure constants:
$$
k_{ab}=-k_{ba}\, ,\qquad  f_{ab}{^c}=-f_{ba}{^c}\, ,\qquad
               V_{ab}^{cd}=V_{ab}^{dc}=-V_{ba}^{cd}\, ,\quad ...
$$
The algebra $(1)$ is clearly not a Lie algebra, but is often referred
to as ``nonlinear Lie algebra".

The Jacobi identity for the bracket product $(1a)$ implies
$$
{\partial W_{ab}\over\partial T_d}W_{cd} + (abc) = 0\, ,      \eqno(1b)
$$
where $(abc)$ denotes cyclic permutation of $a$, $b$ and $c$.
This is a generalization of the Jacobi identity for the usual Lie
algebras. It implies a set of identities for the structure constants;
in particular, $f_{ab}{^d} f_{cd}{^e} + (abc) = 0$.

The problem of constructing a gauge theory based on the nonlinear
extension of a Lie algebra has been studied by Ikeda and Izawa [1]. They
introduced a set of gauge potentials $A^a{_\m}$ and an additional set
of scalar fields $\phi_a$, with {\it gauge transformations} defined by
$$\eqalign{
&\d_0{A^a}_\m = \part_\m\xi^a + W_{bc}^a{A^b}_\m\xi^c \, , \cr
&\d_0\phi_a =- W_{ab}\xi^b \, . }                              \eqno(2)
$$
Here, $W_{ab}$ is a function of $\phi_a$, $\, W_{ab}=W_{ab}(\phi)$, and
we introduce the notation
$$
W_{ab}^c\equiv {\part W_{ab}\over\part\phi_c}\, ,\qquad
W_{ab}^{cd}\equiv {\part^2 W_{ab}\over\part\phi_c\part\phi_d} \, .
$$
These transformations satisfy the following {\it gauge algebra\/}:
$$\eqalign{
&[\d_0(\xi_1), \d_0(\xi_2)] A^a{_\m}=\d (\xi_3) A^a{_\m} -
               \xi^c_1\xi^d_2 W^{ab}_{cd} D_\m\phi_b \, ,  \cr
&[\d_0(\xi_1), \d_0(\xi_2)] \phi_a=\d (\xi_3)\phi_a \, , }  \eqno(3)
$$
where $\,\xi^a_3= W_{bc}^a\xi^b_1\xi^c_2$,  and
$\,D_\m\phi_b=\part_\m\phi_b+ W_{bc}{A^c}_\m$ is the covariant derivative
of $\phi_b$. If the above gauge transformations are to represent some
Lagrangian symmetry, they should be closed at least on shell, i.e. the
relation  $D_\m\phi_b=0$ should be an equation of motion.

The algebra (3) of the gauge transformations is based on the nonlinear
algebra (1) in the sense that the functions $W_{ab}(\phi)$ in (3)
are antisymmetric and satisfy the identities
$(\partial W_{ab}/\partial\phi_d) W_{cd} + (abc)=0$, in complete
agreement with the corresponding Jacobi identities for $W_{ab}(T)$.

By considering the commutator of two covariant derivatives on $\vphi_b$
one defines the curvature
$\,{R^a}_{\m\n}\equiv\part_\m{A^a}_\n - \part _\n{A^a}_\m +
                                          W_{bc}^a{A^b}_\m{A^c}_\n$.
Since the curvature does not transforms homogeneously, there is no
standard prescription for constructing an action invariant
under these gauge transformations. Ikeda and Izawa [1-3] found that
{\it the classical action}  $I_0=\int d^2x\cL$ is determined by
$$
\cL=-{\fr 1 2}\e^{\m\n}\bigl[ ( {R^a}_{\m\n}\phi_a
        +(W_{ab}-W_{ab}^c\phi_c){A^a}_\m{A^b}_\n \bigr]  \, . \eqno(4)
$$
The gauge invariance follows from
$\d\cL=-\part_\m\bigl[ \ve^{\m\n}(W_{ab} - W_{ab}^c\phi_c)
                                               {A^a}_\nu\xi^b \bigr]$.
Moreover, the equations of motion for ${A^a}_\m$ and $\phi_a$ are given
by
$$\eqalign{
&{F_a}^\m \equiv -\e ^{\m\n} D_\n\phi_a = 0\, ,\cr
&F^a \equiv -{\fr 1 2}\e ^{\m\n}{R^a}_{\m\n} = 0\, ,}         \eqno(5)
$$
and, as a consequence, the gauge algebra $(3)$ is closed on shell.

Let us now introduce a convenient notation
$$
\vphi^i=({A^a}_\m,\phi_b) \, ,\qquad \xi^\a=(\xi^a) \, ,
$$
and observe that the gauge transformations (2) can be rewritten in
the form
$$
\d_0\vphi^i=\xi^\a T^i_\a (\vphi) \, .                        \eqno(6)
$$
Gauge invariance of the classical action implies Noether identities
$T^i_\a F_i=0$, where $F_i$ are the classical field equations,
$F_i\equiv \d I_0/\d\vphi^i=({F_a}^\m,F^b)$.

The structure of a classical theory with local symmetries is greatly
clarified by the properties of its gauge algebra. Although the most
natural framework for studying classical gauge algebra is the
hamiltonian approach, the related informations can also be obtained
within the Lagrangian formalism.
Since the change of a functional $I[\vphi]$ under the gauge
transformations (6) has the form $\d_0 I[\vphi]=\xi^\a T^i_\a\d_i
I[\vphi]$, we can introduce the Lagrangian generators of $\xi^\a$
transformations by the relation $\d_0\equiv \xi^\a\G_\a$:
$$
\G_\a =T^i_\a\d_i \, .                                        \eqno(7)
$$
The algebra of the generators is, in general, not closed:
$$
[\G_\a,\G_\b]=\bar f_{\a\b}{^\g}\G_\g+E_{\a\b}^{ij}{F_j}{\d_i}\, .\eqno(8a)
$$
The explicit content of this relation in our case can be found by
rewritting Eq.(3) in the form
$$
\bigl[\d_0(\xi_1), \d_0(\xi_2)\bigr] =\d (\xi_3) +
  \xi_1^c\xi_2^d W^{ab}_{cd} \ve_{\m\n}{F_b}^\n {\d \over \d{A^a}_\m} \, ,
$$
whereupon one easily finds
$$
[\G_\a ,\G_\b]=W_{\a\b}^\g \G_\g +
      W^{cd}_{\a\b}\ve_{\m\n}{F_d}^\n{\d \over \d {A^c}_\m}\, .  \eqno(8b)
$$
We see that the coefficients $\bar f$  and $E$ of the gauge
algebra $(8a)$ are not constants but depend on the fields $\phi_a$.

The commutators (8) satisfy certain consistency requirements following
from the related Jacobi identities. These requirements in general lead
to a natural introduction of new structure functions [8]. However,
explicit calculation shows that these structure functions vanish in
this case.

Thus, the set of structure functions $(T,\bar f,E)$ represents a complet
description of the classical gauge structure of the theory (4), based
on the nonlinear algebra (1).

\subsection{3. Generalized BRST quantization} 

We shall now use a generalized BRST method [7,8] to quantize the theory
(4), whose gauge symmetry is determined by Eqs.(2) and (8b). It
represents a generalization of the BV quantization procedure [6], and
provides an off-shell nilpotent BRST symmetry of the gauge-fixed action.
\subsub{BRST transformations.} The first step in this approach is
the construction of the BRST transformations. Let us start by
introducing for each gauge parameter $\xi^\a$ a ghost $c^\a$; then, to
each field $\Phi^A=(\vphi^i,c^\a)$ we associate the antifield
$\Pi_A=(\vphi_i^*,c_\a^*)$. The Grassmann parities $(\ve)$ and the
ghost numbers of these variables are given in Table 1, where $\ve_i
=\ve(\vphi^i)$ and $\ve_\a =\ve(\xi^\a)$.

Following Batalin and Vilkovisky [6] we define the BRST transformation
$s$ as
$$
s X =(S,X) \, ,                                               \eqno(9)
$$
where $(X,Y)$ is the antibracket of $X$ and $Y$,
$$
(X,Y)={\part_R X\over\part \Phi^A}{\part_L Y\over\part\Pi_A}
      -{\part_R X\over\part \Pi_A}{\part_L Y\over\part\Phi^A}\, ,
$$
and the BRST generator $S=S(\Phi^A,\Pi_A)$ is the solution of the
master equation $(S,S)=0$. The master equation is usually solved by
expanding $S$ in a number of antifields, $S=S_0+S_1+S_2+\cdots$, and
using $S=S_0 -c^\a T^i_\a\Pi_i +\cdots$ as the boundary condition,
with $S_0=I_0$. The off-shell nilpotency of the BRST transformations
follows from $(S,S)=0$.

Following the ideas of the Hamiltonian BRST formalism, we shall try to
find the BRST generator by adding to $S_0$ all possible terms containing
classical structure functions combined with $\Phi^A$ and $\Pi_A$ so
that gh$(S)=0$, while the coefficients of the various terms are
determined by the master equation. Limiting our discussion to dynamical
systems characterized by the structure functions $(T,\bar f,E)$ we
obtain the result [8]
$$\eqalign{
&S=S_0 + S_1 + S_2 \, ,\cr
&S_1=-c^\a T^i_\a \Pi_i +{\fr 1 2}(-)^{\b+1}c^\b c^\a
                                  \bar f^\g_{\a\b}\,\Pi_\g \, ,\cr
&S_2={\fr 1 4}(-)^{i+\b+1}c^\b c^\a E_{\a\b}^{ij}\,\Pi_j\Pi_i\, .}\eqno(10)
$$
\noindent This compact form of $S$ shows very clearly the connection of
the BRST structure to the classical gauge algebra.

After introducing the component notation,
$$
c^\a=(c^a), \qquad \Pi_A=(A_a^{*\m},\phi^{*a},c_a^*)\, ,
$$
a direct calculation based on Eq.$(11)$ leads to the following
component expression for $S$:
$$\eqalign{
&S_1 = -(D_\m c^a) A_a^{*\m} +W_{ab}\,c^b\phi^{*a}
       -{\fr 1 2}W_{ab}^c\,c^bc^ac_c^* \, ,\cr
&S_2=-{\fr 1 4}W^{ab}_{cd}\ve_{\m\n}\,c^d c^c A_b^{*\n}A_a^{*\m}\, ,}
                                                              \eqno(11)
$$
where $D_\m c^a\equiv \part_\m c^a - W_{cb}^a{A^b}_\m c^c$.
By using the above result for the BRST generator one can easily find
the BRST transformations of all the fields,
$$\eqalign{
&s{A^a}_\m = D_\m c^a+{\fr 1 2}W^{ab}_{cd}\ve_{\m\n}\,c^dc^c A_b^{*\n}\, ,\cr
&s\phi_a   = -W_{ab}\,c^b \, , \cr
&sc^a      = {\fr 1 2}W_{bc}^a\,c^cc^b  \, .    }             \eqno(12a)
$$
The transformations of the antifields are
$$
sA_a^{*\m} = {F_a}^\m - W_{ab}^c \, c^b A_c^{*\m}\, ,         \eqno(12b)
$$
and similarly for $\phi^{*a}$ and $c_a^*$.
\subsub{Quantum theory.} The classical action (4) is not BRST
invariant, as the gauge algebra is open. Our next step in the
quantization procedure will be to find a BRST\---invariant extension of
$I_0$. The BRST invariant action is not unique: the BV choice is
$I_{BRST}=S$, while we choose
$$
I_{BRST}=I_0-S_2 \, ,                                         \eqno(13)
$$
which is more convenient at the level of gauge-fixed theory, as we
shall see.

It is important to note that the action (13) is {\it degenerate\/.}
A complete understanding of this degeneracy is of central
importance for the construction of the quantum, gauge-fixed theory.
Since in our approach the antifields $\Pi$ will not be eliminated from
the quantum action as in the BV formalism, we first note that $S_2$ is
{\it degenerate with respect to $\Pi$--variables\/}. In fact, we see
from Eq.(11) that $S_2$ is a function of only $A_a^{*\m}$. The
degeneracy of $S_2$ with respect to the sector $(\phi^{*a},c_a^*)$ can
be removed by fixing these components to zero.

The transition to the restricted set ($A_a^{*\m}$) of antifields
resolves the problem of $\Pi$--degeneracy. One should also show
that this restriction is {\it consistent\/}, i.e. that the new set of
variables  $(\Phi^A,A_a^{*\m})$ carries the representation of the
off-shell nilpotent BRST transformations. This consistency follows from
Eqs. $(12a)$ and $(12b)$: the restricted set of variables is seen to be
closed under the off-shell nilpotent BRST transformations from the very
beginning, as the transformations of this set are decoupled from the
transformations of the removed set $(\phi^{*a},c_a^*)$.

After removing the $\Pi$--degeneracy, there remains the usual degeneracy
related to the classical gauge symmetry. The quantum action can be
defined by introducing the gauge breaking term:
$$
I_q=I_{BRST}-s\Psi \, ,                                       \eqno(14a)
$$
where $\Psi$ is the so-called gauge fermion, $gh(\Psi)=-1, \ve(\Psi)=1$
[6], which satisfies certain regularity conditions. To construct
$\Psi$ we introduce additional fields: antighosts  $\bar c_a$ and
multipliers $B_a$. Their Grassmann parities and ghost numbers are
defined as
$\,\ve(\bar c_a)=1,\,\,gh(\bar c_a)=-1,\,\,\ve(B_a)=0,\,\, gh(B_a)=0,\,$
and their BRST transformations are
$$
s\bar c_a=B_a\, ,\qquad sB_a=0\, .
$$
We can choose $\Psi$ to be a function on the restricted set
$(\Phi^A,A_a^{*\m},\bar c_a,B_a)$. The simple choice
$\Psi=\Psi(\Phi^A,\bar c_a)$ leads to
$$
s\Psi =(s\Phi^A){\part_L\Psi\over\part\Phi^A} +
   B_a{\part_L\Psi\over\part\bar c_a}\equiv I_{FP}+I_{GF}\, . \eqno(14b)
$$

The complete quantum theory is now determined by the generating functional
$$\eqalign{
&Z_\Psi=\int D\m \exp[i(I_{BRST}-s\Psi)]  \, ,\cr
&D\m=D\Phi^A DA_a^{*\m} D\bar c_a DB_a \, .   }               \eqno(15)
$$
Here, the measure is defined only over the independent variables, and
$Z_\Psi$ is nondegenerate by construction. Note that the quantum action
$I_q$ is invariant under the off-shell nilpotent BRST transformations
(this implies $\Psi$--independence of $Z_\Psi$ and, consequently, gauge
invariance of the $S$--matrix). The condition $I_q(\Pi=0)=I_0$ ensures
the correct classical limit of the theory.

The simplest choice for $\Psi$ is given by a bilinear combination of
antighosts times gauge conditions (linear gauges):
$$
\Psi=\bar c_a {M^a}_i\vphi^i \, ,
$$
where $\chi^a= {M^a}_i\vphi^i$ are gauge conditions defined on the
classical fields $\vphi^i$, and $M$ is a field--independent matrix.
The structure of $M$ is determined by the regularity conditions imposed
on $\Psi$.
\subsub{Comparisson to the BV formalism.} It is instructive to compare
our approach to the BV method [6]. The basic difference lies in the
treatment of antifields.

In the BV approach the quantum action is obtained by the replacing
$\Pi_A\to {\part\Psi/\part\Phi^A}$ in S:
$$
I_{BV} = S' - B_a{\part_L\Psi\over\part\bar c_a} \, , \qquad
         S'\equiv  S(\Phi^A,\Pi_A= {\part_L\Psi/\part\Phi^A})\, .
$$
We shall show that the integration over the antifields $A_a^{*\m}$ in
Eq.(15) yields an effective action that coincides with the BV result.
To see that we introduce the notation $S_1=\L^A\Pi_A$,
$S_2=\L^{AB}\Pi_B\Pi_A$, and observe that the quantum action (14) can
be written in the form
$$
I_q=I_0+\L^A{\part_L\Psi\over\part\Phi^A}
  +\L^{AB} {\part_L\Psi\over\part\Phi^B}{\part_L\Psi\over\part\Phi^A}
  -B_a{\part\Psi\over\part\bar c_a}-\D =I_{BV} - \D \, ,
$$
where
$$\eqalign{
\D &\equiv \L^{AB}(\Pi_B-{\part_L\Psi\over\part\Phi^B})
                 (\Pi_A-{\part_L\Psi\over\part\Phi^A})    \cr
   &=-{\fr 1 4}W^{cd}_{ab}\e_{\m\n}\, c^bc^a
     \left( A_d^{*\n}-{\part_L\Psi\over\part{A^d}_\n}\right)
     \left( A_c^{*\m}-{\part_L\Psi\over\part{A^c}_\m}\right) \, .  }
$$
It is now evident that the integration over $A_c^{*\m}$ in $Z_\Psi$
eliminates the term $\D$, and the resulting effective quantum action
coincides with the BV expression, $I'_q=I_{BV}$.

The BRST transformations obtained in the BV formalism after fixing the
gauge are nilpotent only on shell. Indeed, from the general relation
$s'\Phi^A=(S,\Phi^A)\ver{2}_{\Pi=\part\Psi/\part\Phi}$ we have
$$
s'^2{A^a}_\m = -{\fr 1 2}W^{ab}_{cd}\ve_{\m\n}\, c^dc^c \cF_b{^\n} \, ,
$$
where $\cF_b{^\n}$ is the equation of motion for $A^b{_\n}$ following
from $I_{BV}$:
$$
\cF_b{^\n}=F_b{^\n}-W_{bc}^d\, c^c{\part\Psi\over\part A^d{_\n}}
                  -B_c{\part^2\Psi\over\part A^b{_\n}\part\bar c_c} \, .
$$
Thus, $s'^2=0$ only on shell, as a consequence of the nonclosure of
the classical gauge algebra $(W^{ab}_{cd}\ne 0)$. BRST transformations
of other variables are nilpotent off shell.

The essential role of the antifields $A_b^{*\n}$ in the quantum
action (14) is to ensures the off--shell nilpotency of the BRST
transformations in the gauge--fixed theory.

\subsection{4. Gauge theory based on quadratically extended
            Poincar\' e algebra}       

Attempts to formulate the theory of gravity as a gauge theory led to
considering the Poincar\'e gauge theory as a candidate for a consistent
theory of gravity. We shall consider here, as an application of the
previous general formalism, the quantization of two--dimensional gauge
theory based on quadratically extended Poincar\'e algebra.
\subsub{Poincar\'e gauge theory.} The basic dynamical variables of
this theory in 2d are the diad $b^a{_\m}$ and the connection
$A^{ab}{_\m}$, associated with the translation and Lorenz subgroup of
the Poincar\'e group, respectively. Here, $a,b,...=0,1$ are the local
Lorenz indeces, while $\m,\n,...=0,1$ are the coordinate indeces.
The structure of the Poincar\'e group is also reflected in the existence
of two kinds of gauge field strengths: the torsion  $T^a{_{\m\n}}$, and
the curvature $R^{ab}{_{\m\n}}$. The most general action of the
Poincar\'e gauge theory in 2d, which is at most quadratic in gauge field
strengths, has the form [9,10]:
$$
I'_0=\int d^2x\,b\left( {1\over 16\a}R^{ab}{_{\m\n}}R_{ab}{^{\m\n}}
        -{1\over 8\b}T^a{_{\m\n}}T_a{^{\m\n}}-\g\right) \, ,  \eqno(16)
$$
where $b=\det(b^a{_\m})$, and $\a,\b$, $\g$ are constants.
In 2d the Lorenz connection $A^{ab}{_\m}$ can be parametrized as
$A^{ab}{_\m}=\ve^{ab}A_\m$, so that
$$\eqalign{
&R^{ab}{_{\m\n}}=\ve^{ab}R_{\m\n}\, ,\qquad
                 R_{\m\n}\equiv\part_\m A_\n -\part_\n A_\m \, ,\cr
&T^a{_{\m\n}}=D_\m b^a{_\n}-D_\n b^a{_\m} \, ,}
$$
and $D_\m b^a{_\n}=\part_\m b^a{_\n} +\ve^a{_c}A_\m b^c{_\n}$ is the
covariant derivative of the diad field. The action (16) is invariant
under the local Poincar\'e transformations with parameters $\o\ve^{ab}$
and $a^\l$:
$$\eqalign{
&\d_0 b^a{_\m}=\o\ve^a{_c}b^c{_\m}-a^\l{_{,\m}}b^a{_\l}
                                 -a^\l\part_\l b^a{_\m} \, ,\cr
&\d_0A_\m =  -\part_\m\o -a^\l{_{,\m}}A_\l -a^\l\part_\l A_\m \, ,\cr
&\d_0\vphi = -a^\l\part_\l \vphi \, ,\qquad\quad
\d_0\vphi^a=\o\ve^a{_c}\vphi^c -a^\l\part_\l \vphi \, . }
$$
\subsub{Nonlinear extension.} It has been shown [1,3] that the theory
$(16)$ in the region $b\ne 0$ is equivalent to a gauge theory based on
the following quadratic extension of the Poincar\'e algebra:
$$\eqalign{
&[M,M]=0\, ,\qquad [M,P_a]=-\ve_a{^b}P_b \, ,\cr
&[P_a,P_b]=\ve_{ab}(\a M^2+\b\eta^{ab}P_aP_b+\g) \, .}        \eqno(17)
$$
Indeed, by using the general procedure of the previous section with
$A^a{_\m}\to (b^a{_\m},A_\m)$ and $\phi_a\to (\vphi,\vphi_a)$, one
obtains the gauge invariant action in the form
$$
I_0 = \int d^2x\left[\,{\fr 1 2}\ve^{\m\n}(\vphi R_{\m\n} +
\vphi_a T^a{_{\m\n}}) -b(\a\vphi^2 +\b\vphi_a\vphi^a +\g)\,\right]\, .
                                                              \eqno(18)
$$
This expression represents the first--order formulation of the action
$(16)$, with $(\vphi,\vphi_a)$ playing the role of auxiliary fields.
It is equivalent to $(16)$ in the region $b\ne 0$, in the sense that
(18) reduces to (16) when the auxiliary fields are eliminated.
In the weak coupling limit, $\a,\b,\g\to 0$, the theory (18) becomes
the topological $ISO(1,1)$ gauge theory [11]. The clarification of the
dynamical content of the theory (16) [10] helps us to understand this
relationship more clearly.

The equations of motion for ${b^a}_\m$, $A_\m$, $\vphi$ and $\vphi_a$
are given by
$$\eqalign{
&{F_a}^\m =\ve^{\m\n} D_\n\vphi_a
            -b{h_a}^\m(\a\vphi^2 +\b\vphi_a\vphi^a +\g)=0 \, ,   \cr
&F^\m = \ve^{\m\n}(\part_\n\vphi +\e^{ab}\vphi_a b_{b\n})=0\, ,\cr
&F = {\fr 1 2}\ve^{\m\n}R_{\m\n}-2\a b\vphi = 0   \, ,         \cr
&F^a = {\fr 1 2}\ve^{\m\n}T^a{_{\m\n}} - 2\b b\vphi^a =0 \, ,}
                                                              \eqno(19)
$$
where ${h_a}^\m$ is the inverse diad field and
$D_\n\vphi_a=\part_\n\vphi_a -\e^c{_a}A_\n\vphi_c$.

The action is invariant under the gauge transformatios with
parameters $(\xi,\xi^a)$:
$$\eqalign{
&\d_0 A_\m =\part_\m\xi +2\a\ve_{bc}\xi^b{b^c}_\m\vphi \, ,\cr
&\d_0 {b^a}_\m = \part_\m\xi^a +\ve^{ab}\bigl( -\xi b_{b\m}
                +\xi_b A_\m\bigr) + 2\b\ve_{bc}\xi^b{b^c}_\m\vphi^a\, ,\cr
&\d_0\vphi   = \ve_{ab}\xi^a\vphi^b \, ,  \qquad
\d_0\vphi_a = \ve_{ab}\bigl[-\xi\vphi^b
             +\xi^b(\a\vphi^2 +\b\vphi_c\vphi^c +\g )\bigr]\, . } \eqno(20)
$$
After redefining the parameters by
$\xi\to -\o -a^\l A_\l$, $\xi^a\to -a^\l b^a{_\l}$,
one easily finds that these transformations reduce, up to the equations
of motion, to the standard local Poincar\'e transformations.

Now, we introduce the gauge generators $(\G,\G_a)$ corresponding to the
$(\xi,\xi^a)$ transformations by $\d_0=\xi\G +\xi^a\G_a$. The algebra
of the generators is closed only {\it on shell}:
$$\eqalign{
&[\G,\G]=0 \, , \qquad [\G,\G_b] = -\ve_b{^c}\G_c \, ,\cr
&[\G_a,\G_b]=-2\ve_{ab}\bigl( \a\vphi\G +\b\vphi^c\G_c\bigr)
           +2\ve_{ab}\ve_{\m\n}\left( \a F^\n{\d\over\d A_\m}
           +\b F^{c\n}{\d\over\d{b^c}_\m}\right) \, . }      \eqno(21)
$$
Equations (20) and (21) determine the structure functions
$(T,\bar f,E)$ of the classical gauge algebra.
\subsub{Quantization.} Let us now apply the general method developed in
the previous section to study the quantization of the theory (18). The
structure of classical fields, ghosts and antifields is given in Table 3.

The evaluation of the BRST generator on the basis of Eq.(11) leads to
the result
$$\eqalign{
S_1 =&-(\part_\m c +2\a\ve_{bc}c^b{b^c}_\m\vphi )A^{*\m}        \cr
     &-\bigl[\part_\m c^a + \ve^{ab}(-cb_{b\m} + c_b A_\m)
         +2\b\ve_{bc}c^b{b^c}_\m\vphi^a\bigr]{b^*_a}^\m         \cr
     &-\ve_{ab} c^a\vphi^b\vphi^* -\ve_{ab}\bigl[ -c\vphi^b
         +c^b(\a\vphi^2 +\b\vphi_c\vphi^c +\g )]\vphi^{*a}     \cr
     & -\ve_{ab}\bigl[ \a\vphi c^ac^bc^*
        + (\b c^ac^b\vphi^c - c^a c\eta^{bc})c^*_c\bigr]  \, , \cr
S_2 =&{\fr 1 2} \ve_{ab}\ve_{\m\n}c^ac^b (\a A^{*\n}A^{*\m}
        +\b b^{*d\n}{b^*_d}^\m ) \, .  }                      \eqno(22)
$$
{}From here one can easily find the form of the BRST transformations on
the fields,
$$\eqalign{
&sA_\m= \part_\m c +2\a\ve_{bc}c^b{b^c}_\m\vphi
                            -\a\ve_{bc}c^bc^c\ve_{\m\n}A^{*\n} \, ,\cr
&sb^a{_\m}=\part_\m c^a + \ve^{ab}(-cb_{b\m} + c_b A_\m)
       +\b\e_{bc}(2c^b{b^c}_\m\vphi^a-c^bc^c\ve_{\m\n}b^{*a\n})\, ,\cr
&s\vphi=\ve_{ab} c^a\vphi^b \, ,\qquad\quad
s\vphi_a=\ve_{ab}\bigl[ -c\vphi^b
                         +c^b(\a\vphi^2 +\b\vphi_c\vphi^c +\g )]\, ,\cr
&sc= \a\ve_{ab}\vphi c^a c^b \, ,\qquad\,
sc^a=\b\ve_{bc}c^bc^c\vphi^a +\ve^a{_b}c^b c \, ,}           \eqno(23a)
$$
and the antifields,
$$\eqalign{
&sA^{*\m}=F^\m-\ve^{ab}c_b b_a^{*\m} \, ,\cr
&sb_a^{*\m}=F_a{^\m}+\ve_{ab} \bigl( 2\a c^b\vphi A^{*\m}-cb^{*b\m}
            + 2\b c^b\vphi^c b_c^{*\m}\bigr) \, , }           \eqno(23b)
$$
and similarly for other antifields. It should be observed that the set
of all fields together with the restricted set of antifields
$(A^{*\m},b_a^{*\m})$ also carries the representation of the BRST
transformations: these variables transform into each other under $s$
and, moreover, $s^2=0$ off--shell.

The BRST invariant action (13) is degenerate in the antifield sector, as
$S_2$ contains only $A^*_\m$ and $b_a^{*\m}$. This degeneracy can be
removed by imposing the following extra conditions:
$\vphi^*=\vphi^{*a}=c^*=c^{*a}=0$.  The restricted set of antifields
$(A^*_\m,b_a^{*\m})$ is sufficient to define the representation of
the off-shell nilpotent BRST transformations, so that the whole
procedure is completly consistent.

The quantum action is now determined by the expression $(14a)$.
The final form of the quantum theory depends on the choice of gauge.
The authors of Ref. [1] considered two gauge choices: the temporal
gauge $(A_0=0,b^a{_0}=0)$, and the background--covariant gauge
$[\vphi_a=0, \part_\m(\tilde g^{\m\n}A_\n)=0].$ The related gauge
fermions are given by the expressions
$a)\,\Psi_1 = \bar c A_0 + \bar c_a{b^a}_0$ and
$b)\,\Psi_2 = \bar c^a\vphi_a +
                         \bar c\part_\m\,(\tilde g^{\m\n} A_\n) $,
while the BRST transformations of antighosts and multipliers are of the
standard forms. The above gauge choices, in conjunction with the
equations of motion, imply that the diad field is degenerate, i.e.
$b=0$. This is acceptable if we interpret the action (18) as describing a
Yang--Mills theory. However, the correct quantization of the
gravitational theory (16) requires a gauge condition consistent with
$b\ne 0$. $c)$ The conformal gauge $b^a{_\a}=e^\s \d^a_\b$ was used
in [9] for classical calculations, but its nonlinearity makes it not so
convenient in quantum theory [12]. $d)$ The Landau--type gauge
$(\part^\m A_\m=0, \part^\m b^a{_\m}=0)$, and $e)$ the light cone gauge
$(A_+=0,b^+{_+}=b^-{_+}=0)$ were used in Ref. [12] in considerations
related to the question of renormalizability of $R^2+T^2$ theory of
gravity.

After choosing the gauge conditions, the quantum, nonlinear Poincar\'e
gauge theory is defined by the generating functional
$$\eqalign{
&Z_\Psi=\int D\m \exp[i(I_{BRST}-s\Psi)]  \, ,\cr
&D\m=Db^a{_\m}DA_\m D\vphi_aD\vphi Dc Dc^a
                  Db_a^{*\m} DA^{*\m} D\bar c DB \, .   }     \eqno(24)
$$
The structure of antighosts $\bar c$ and multipliers $B$ depends on the
choice of gauge.

\subsection{5. Concluding remarks} 

In this paper we have studied the covariant quantization of $2d$ gauge
theories based on nonlinear extension of Lie algebras, in the
generalized Lagrangian formalism. We first constructed the BRST
invariant action $I_{BRST}$ by using the information on the classical
gauge structure of the theory. The BRST symmetry of the gauge--fixed,
quantum theory is off--shell nilpotent. It is realized on the set of
variables $(A^a{_\m},\phi_a,c^a;\,A_a^{*\m})$ containing the restricted
set of antifields $A_a^{*\m}$ as auxiliary variables with nonvanishing
ghost number. The relation of our approach to the BV one is clarified.

The general results are then applied to study the covariant
quantization of a specific nonlinear gauge theory in $2d$, based on the
quadratic extension of the Poincar\'e algebra. This theory is of
particular interest for investigations of the quantum structure of
gravity, as it represents an interesting connection between several
possible formulations  of $2d$ gravity.

The $W_3$ gravity and the dilaton gravity can be treated in a similar way.

\vfill\eject

\subsection{References}                

\item{1.} N. Ikeda and K.-I. Izawa, Prog. Theor. Phys. {\bf 89} (1993) 1077;
                                    Prog. Theor. Phys. {\bf 90} (1993) 223.

\item{2.} N. Ikeda and K.-I. Izawa, Prog. Theor. Phys. {\bf 90} (1993) 237.

\item{3.} N. Ikeda, Kyoto preprint RIMS--953 (1993).

\item{4.} K. Shoutens, A. Sevrin and P. van Nieuwenhuizen, Phys. Lett.
{\bf B255} (1991) 549; Int. J. Mod. Phys. {\bf A6} (1991) 2891.

\item{5.} \" O. F. Dayi, preprint "BV and BFV Formulation of a gauge
Theory of Quadratic Lie Algebras in $2d$ and a construction of $W_3$
Topological Gravity" (1994).

\item{6.} I. A. Batalin and G. A. Vilkovisky, Phys. Rev. {\bf D28}
(1983) 2567; Phys. Lett. {\bf B120} (1983) 166.

\item{7.} M. Blagojevi\'c and B. Sazdovi\'c, Phys. Lett. {B223} (1989)
325, 331.

\item{8.} M. Blagojevi\'c, B. Sazdovi\'c and M. Vasili\'c, Phys.
Lett.{\bf B236} (1990) 424; Nucl.Phys. {\bf B365} (1991) 467;
\item{} \v Z. Antunovi\'c and M. Blagojevi\'c, Nucl. Phys. {\bf B363}
(1991) 622;
\item{} M. Blagojevi\'c, B. Sazdovi\'c and T. Vuka\v sinac, Mod. Phys.
Lett. {\bf A8} (1993) 349;
\item{} \v Z. Antunovi\'c, M. Blagojevi\'c and T. Vuka\v sinac, Mod.
Phys. Lett. {\bf A8} (1993) 1983.

\item{9.} M. O.Katanaev and I. V. Volovich, Phys. Lett. {\bf B175}
(1986) 413; Ann. of Phys. {\bf 197} (1990) 1;

\item{} I. V. Volovich, Two-dimensional gravity with dynamical torsion
and strings, preprint CERN-TH 4771/87 (1987);

\item{} M. O. Katanaev, Theor. Math. Phys. {\bf 80} (1989) 838; J.
Math. Phys. {\bf 31} (1990) 882; {\bf 32} (1991) 2483.

\item{10.} K. G. Akdeniz, \" O. F. Dayi and A. Kizilers\" u, Mod.
Phys.Lett {\bf A7} (1992) 1757;

\item{} H. Grosse, W.Kummer, P. Pre\v snajder and D. J. Schwarz, J.
Math. Phys. {\bf 33} (1992) 3892;
\item{} T. Strobl, Int. J. Mod. Phys. {\bf A8} (1993) 1383;
\item{} T. Schaller and  T. Strobl, Class. Quant. Grav. {\bf 11} (1994)
331;
\item{} F. Heider and W. Kummer, Preprint TUW-92-15.

\item{11.} T. Fukujama and K. Kamimura, Phys. Lett. {\bf B160} (1985)
259;
\item{} K. Isler and C. A. Trugenberger, Phys. Rev. Lett. {\bf 63}
(1989) 834;
\item{} A. H. Chamseddine and D. Whyler, Phys. Lett. {\bf B228} (1989)
75; Nucl. Phys. {\bf B340} (1990) 595.

\item{12.} W. Kummer and D. J. Schwarz, Phys. Rev. {\bf D45} (1992)
3628; Nucl. Phys. {\bf B382} (1992) 171.

\vfill\eject

\phantom{a}
\vskip1cm

\vskip.3cm
\centerline{{\eightrm Table 1. The Grassmann parities and the ghost
                      numbers of the fields and antifields}}
\vskip-6pt
$$
\vbox{\halign{\hskip2pt #\hfil &&\hskip27pt #\hfil\cr
\noalign{\hrule}\noalign{\smallskip}
       &$\vphi^i$ &$\hskip9pt c^\a$ &\hskip9pt$\Pi_i$ &$\Pi_\a$  \cr
\noalign{\smallskip}\noalign{\hrule}\noalign{\smallskip}
 $\ve$ &$\ve_i$   &$\ve_\a +1$      & $\ve_i +1$      &$\ve_\a$  \cr
 $gh$  &$0$       &\hskip12pt$1$    &\hskip9pt$-1$    &$ -2$     \cr
\noalign{\smallskip}\noalign{\hrule}  } }
$$
\vskip2cm

\vskip.3cm
\centerline{{\eightrm Table 2. The components of the fields and antifields}}
\vskip-6pt
$$
\vbox{\halign{\hskip2pt #\hfil &&\hskip22pt #\hfil\cr
\noalign{\hrule}\noalign{\smallskip}
&${A^a}_\m$ &$\phi_a$ &$c^a$ &$A_a^{*\m}$ &$\phi^{*a}$ &$c_a^*$    \cr
\noalign{\smallskip}\noalign{\hrule}\noalign{\smallskip}
$\ve$ &$0$  &$0$      &$1$    &\hskip7pt$1$  &\hskip7pt$1$ &\hskip7pt$0$ \cr
$gh$  &$0$  &$0$      &$1$    &$-1$          &$-1$         &$-2$        \cr
\noalign{\smallskip}\noalign{\hrule}  } }
$$
\vskip2cm

\vskip.3cm
\centerline{{\eightrm Table 3. The fields and antifields for
                      quadratically nonlinear Poincar\'e gauge theory}}
\vskip-15pt
$$
\vbox{\halign{\hskip2pt #\hfil &&\hskip14pt #\hfil\cr
\noalign{\hrule}\noalign{\smallskip}
&${b^a}_\m$   &$A_\m$    &$\vphi$   &$\vphi_a$    &$c$   &$c^a$
&${b^*_a}^\m$ &$A^{*\m}$ &$\vphi^*$ &$\vphi^{*a}$ &$c^*$ &$c^*_a$   \cr
\noalign{\smallskip}\noalign{\hrule}\noalign{\smallskip}
$\ve$  &0 &0 &0 &0  &1 &1 &1 &1   &1 &1 &0 &0                       \cr
$gh$   &0 &0 &0 &0  &1 &1 &\hskip-7pt$-1$ &\hskip-7pt$-1$
    &\hskip-7pt$-1$ &\hskip-7pt$-1$ &\hskip-7pt$-2$&\hskip-7pt$-2$  \cr
\noalign{\smallskip}\noalign{\hrule}  } }
$$

\end